
\vskip-2.5cm
\line{\hbox{\hskip12.5cm THU-92/15\hfil}}
\line{\hbox{\hskip12.5cm June 1992\hfil}}
\vskip1cm
\def\nn{{\ | \!\!\! |\ }}
\def\vek#1{{\vec #1}}
\def\real{{I\!\! R}}
\def\Tr{{\rm Tr}}
\centerline{{\bf TOPOLOGY OF THE YANG-MILLS CONFIGURATION SPACE}
\footnote\dag{Invited talk to appear in the proceedings of
the International Symposium
on Advanced Topics of Quantum Physics (ISATQP), June 11-16, Shanxi
University, Taiyuan, P.R. China.}}
\vskip1in
\centerline{PIERRE VAN BAAL \parindent=5mm\footnote*{KNAW fellow}}
\centerline{\it Institute for Theoretical Physics,}
\centerline{\it Princetonplein 5, P.O.Box 80.006,}
\centerline{\it NL-3508 TA Utrecht, The Netherlands}
\vskip1.5cm
\centerline{ABSTRACT}
\vskip3mm
{\narrower
It will be described how to uniquely fix the gauge using Coulomb
gauge fixing, avoiding the problem of Gribov copies. The fundamental
modular domain, which represents a one-to-one representation of the
set of gauge invariant degrees of freedom,
is a {\it bounded} convex subset of the transverse gauge fields.
Boundary identifications are the only remnants of the Gribov copies,
and carry all the information about the topology of the Yang-Mills
configuration space. Conversely, the known topology can be shown to
imply that (on a set of measure zero on the boundary) some points
of the boundary coincide with the Gribov horizon.
For the low-lying energies, wavefunctionals
can be shown to spread out ``across'' certain parts of these boundaries.
This is how the topology of Yang-Mills configuration space has an
essential influence on the low-lying spectrum, in a situation where
these non-perturbative effects are {\it not exponentially suppressed}.
\par}
\vfill
\eject
The write-up of my contribution to the International Symposium
on Advanced Topics of Quantum Physics will be a short summary,
with adequate references, where most of the material I have presented
can be found. The observation concerning Henyey's gauge copies
has not been published before.

Gauge fixing has remained an essential ingredient for studying
non-abelian gauge theories~[1], despite the many attempts of finding
gauge invariant variables. This is not too surprising, as the Yang-Mills
configuration space is topologically non-trivial~[2]. There exists no
choice of (affine) coordinates suitable for the whole manifold. Gauge
fixing can be conveniently seen as just one particular choice of
coordinates~[3]. They are only locally defined, and one would need
different coordinate patches and transition functions to describe
the whole manifold~[4]. Trying to extend the coordinate patch to the
whole manifold has as a consequence that the gauge condition, beyond a
certain distance in the configuration space, no longer uniquely fixes
the gauge, as was first discovered by Gribov~[5].
Analysing this in a finite spatial volume
using the Hamiltonian formulation ($A_0=0$ and using the Coulomb
gauge to fix the time-independent gauge parameters) has demonstrated
the usefulness of this approach. In that case the transition functions
were described by topologically non-trivial gauge transformations~[6].
The reason for its success lies in the fact that asymptotic freedom
in four dimensions
guarantees that the effective coupling constant in a small volume is
small~[7], and that non-perturbative effects will only be of importance
for a few low-energy modes.

More recently, a different way of describing the Yang-Mills
configuration space was developed, by using a functional method
to not only satisfy the Coulomb gauge condition $\partial_iA_i=0$,
but to also pick from the different Gribov copies a unique
configuration. The collection of these configurations should thus form a
fundamental modular domain, in other words it should form a one-to-one mapping
with the Yang-Mills configuration space. The relevant functional
is the $L^2$-norm of the gauge field: $\nn A_i\nn^2=\int_M\Tr(A_i^\dagger
A_i)$. For each gauge invariant field configuration this gives a Morse
function on the gauge orbit. One easily verifies that stationary
points of this Morse function satisfy the Coulomb gauge condition
and that the Hessian at the stationary point is precisely given by the
Faddeev-Popov operator, whose determinant measures the volume of the
gauge orbit. This procedure to fix the gauge has been known already
for quite some time~[8], as well as the obvious conclusion that the
absolute minimum provides the natural way of choosing a unique
representative~[9], whose recent rediscovery~[10] has resulted in a
recurrent interest in this problem.

Gribov's original conjecture~[5] was that one could find a unique
representative by demanding the Faddeev-Popov operator to be positive.
This region in the space of connections can be shown to be convex and
is called the Gribov region.
In a finite volume one can show that in each direction of configuration
space, the distance of its boundary to the origin is finite~[11].
However, this would mean that this choice is only unique if the above
discussed Morse function has all its minima to be absolute minima.
The existence of a Gribov horizon, where the Faddeev-Popov determinant
vanishes (we will be reserving the name horizon for the more stringent
condition that the lowest eigenvalue of the Faddeev-Popov operator
vanishes), does imply that points just outside the horizon have
copies just inside, as Gribov showed. But more importantly, taking the
global nature of the Morse function into account, implies
two possibilities. Either
near the horizon the points inside the Gribov region correspond to
local minima, which therefore have to be copies of an absolute minimum,
by definition also inside the Gribov region~[9,10]. This happens if the
third order term of the expansion of the Morse function around the
stationary point, in the direction where the Hessian vanishes, is
non-vanishing. Or, in the case this third order term vanishes,
the (local) minimum bifurcates at the horizon in two local minima
and one saddle point~[12].
As the two local minima in the above bifurcation coalesce at the
horizon, the relevant gauge transformation has to be
topologically trivial. By definition, these
gauge copies lie inside the Gribov region.

Care is required if the gauge configuration is
reducible, as in that case the gauge group does not act freely.
However, the gauge subgroup that leaves the reducible connection
fixed, generally does not support topologically non-trivial maps.
Nevertheless, reducible connections give rise to singularities in
the physical configuration space~[13], where the Morse theory arguments
might not be valid. These singularities turn out to be particularly
treacherous in the case of an abelian connection, which are the simplest
examples of reducible connections. As the Coulomb gauge does not
fix the constant gauge transformations, one has to still divide out
these constant gauge transformations to obtain the fundamental modular
domain.

With this in the back of
our minds let us reconsider an old example due to Henyey~[14], for
gauge copies (inside the Gribov horizon) that are related by a
homotopically trivial gauge transformation. As was explained in
ref.[12], his was an example related to a bifurcation of local
minima at the horizon. Explicitly Henyey's example was given by~[14]
an abelian SU(2) connection on compactified $\real^3$:
$$\eqalignno{
\vek A(\beta)&=a(\beta)\hat\phi\tau_3,&\cr
a(\beta)={1\over 2r\sin\theta}-&{b+r^2\sin^2\theta({d^2b\over dr^2}+
{4\over r}{db\over dr})\over \beta^{-1}\sin(2r\beta b\sin\theta)},&\cr
g(\beta)=\exp(& i\beta r
b\sin\theta (e^{i\phi}\tau_-+e^{-i\phi}\tau_+)).&\cr}
$$
where $(r,\theta,\phi)$ are the spherical coordinates, $\tau_i$ the
Pauli matrices, $b$ an arbitrary function of $r$ and
$g(\pm\beta)$ defines the two gauge copies $[g(\pm\beta)]A(\beta)$,
that will coalesce at $\beta=0$, which is easily verified to coincide with
the Gribov horizon for a suitable choice of the radial function $b$~[12].
Unfortunately, close inspection shows that, since $g(-\beta)=
\tau_3g(\beta)\tau_3$ and since $A(\beta)$ commutes with $\tau_3$, the two
copies $[g(\pm\beta)]A(\beta)$ are related by a constant gauge transformation.
As we still have to divide out these constant gauge transformations,
Henyey's example does not really provide an example of two
gauge copies inside the Gribov horizon, that are related by
a topologically trivial gauge transformation.

Fortunately, however, the general Morse theory argument is sufficiently
strong not to have to worry, that bifurcations of the above type do not
occur. To find an example one simply should look for an
irreducible connection on the Gribov horizon, where the third order term
in the expansion around the stationary point of the Morse function
vanishes in the direction of the zero eigenvector of the Hessian.
In a recent paper, that analysed gauge fields on a three-sphere~[15],
such configurations featured prominently as the so-called sphaleron
modes $(u,v)$. Part of the Gribov horizon is given by the equation
$u+v=1.5$ (for $u$ and $v\in[0,1.5]$) and one easily verifies these
configurations are not reducible and have a Morse function that vanishes
to third order in the direction of the zero eigenvector of the
Faddeev-Popov operator.

We now come back to the issue of the fundamental modular domain. It can
be shown that the collection of absolute minima of the Morse function
is also convex~[10]. One would thus expect it to be contractable, which
is in flagrant contradiction with the assumption that it is a in
one-to-one correspondence with the physical configuration space. The
catch is that the absolute minimum of the Morse function need not be
unique. Using the convexity of the set of absolute minima (also called
$\Lambda$) it is not hard to show~[12] that $\Lambda$ has a boundary, which
if it does not coincide with the Gribov horizon, necessarily corresponds
to a degenerate absolute minimum of its Morse function. The other
absolute minimum is necessarily a gauge copy and also a point on the
boundary. Thus the set of absolute minima will only become a
fundamental modular domain if these appropriate boundary identifications
are taken into account. In this way it will support the known
non-trivial topology of the physical configuration space. On the
other hand, the known non-trivial topology implies the existence of
non-contractable $n$-spheres. Those can only arise through boundary
identifications if all points of a suitable $n-1$-dimensional subspace
of the boundary are identified (gauge equivalent). These points
consequently coincide with the Gribov horizon. These ``singular''
boundary points, however, form a subset of the boundary of zero
measure (which does {\it not} mean they are unimportant). Note
that as the Gribov region is bounded in each direction, the same holds
for the fundamental modular domain.

The dynamical consequence in the context of gauge theories in a finite
volume is that at increasing volume (i.e. increasing coupling constant)
the wavefunctional starts to spread out over configuration space, and
unavoidably will start to overlap with the boundary of the fundamental
modular domain. Important is that the size of the volume controls this
process. At very small volumes the boundary is irrelevant (for a torus
typically below $0.1-0.01~fm^3$), whereas at increasing volume
first the lowest energy modes start to be sensitive to the boundary
identifications. In this way the energy of electric flux and the
low-lying glueball spectrum were calculated in volumes up to $1~fm^3$
in the geometry of a torus~[12,16], which agreed with the lattice
Monte Carlo results obtained in the same volume within the 2\% statistical
error of the latter~[17]. For the torus geometry the zero energy modes
are parametrized by the constant abelian modes, $A_i=C_i\tau_3/(2L)$
(where $L$ is the length of the torus). The Gribov horizon and the
boundary of the fundamental modular domain in $C$-space are given by
cubes, centered at $C=0$ and respectively with sides of length $4\pi$
and $2\pi$~[12]. A recent analysis on the three-sphere is aimed
at going to larger volumes~[15].
\vskip2cm
\noindent{\bf References}
\item{[1]} C.N. Yang and R.L. Mills, Phys. Rev. 96(1954)191
\item{[2]} I. Singer, Comm. Math. Phys. 60(1978)7
\item{[3]} O. Babelon  and C. Viallet, Comm. Math. Phys. 81(1981)515.
\item{[4]} W. Nahm, in: IV Warsaw Symp. on Elem. Part. Phys.,
    ed. Z. Ajduk, p.275 (Warsaw, 1981)
\item{[5]} V. Gribov, Nucl. Phys. B139(1978)1
\item{[6]} P. van Baal, {\it in}: Probalilistic Methods in Quantum Field
Theory and Quantum Gravity, ed. P. Damgaard et al (Plenum, New York,
1990) p.131.
\item{[7]} M. L\"uscher, Phys. Lett. 118B(1982)391; Nucl. Phys. B219(1983)233
\item{[8]} T. Maskawa and H. Nakajima, Prog. Theor. Phys. 60(1978)1526;
K. Wilson, in: Recent developments in Gauge Theories, ed. G.
't Hooft, et al (Plenum Press, New York, 1980) p. 363;
G. 't Hooft, Nucl. Phys. B190[FS3](1981)455
\item{[9]} M.A. Semenov-Tyan-Shanskii and V.A. Franke, Zapiski Nauchnykh
Seminarov\hfill\break Leningradskogo Otdeleniya Matematicheskogo Instituta
im. V.A. Steklov AN SSSR, 120(1982) 159. Translation:
(Plenum Press, New York, 1986) p.1999
\item{[10]} G. Dell'Antonio and D. Zwanziger,
in: Probabilistic Methods in Quantum Field Theory and Quantum
Gravity, ed. P.H. Damgaard et al, (Plenum Press, New York, 1990) p.107;
Comm. Math. Phys. 138(1991)291
\item{[11]} G. Dell`Antonio and D. Zwanziger, Nucl. Phys. B326 (1989)
333; D. Zwanziger, Critical limit of lattice gauge theory,
NYU preprint, December 1991
\item{[12]} P. van Baal, Nucl. Phys. B369(1992)259
\item{[13]} S. Donaldson and P. Kronheimer, {\it The geometry of
four-manifolds} (Oxford University Press, 1990);
D. Freed and K. Uhlenbeck, {\it Instantons and
four-manifolds}, M.S.R.I. publications, Vol 1 (Springer, New York,
1984).
\item{[14]} F.S. Henyey, Phys. Rev. D20(1979)1460
\item{[15]} P. van Baal and N.D. Hari Dass, ``The theta dependence
beyond steepest descent'', Utrecht preprint, THU-92/03, January 1992,
to appear in Nucl. Phys. B.
\item{[16]} J. Koller and P. van Baal, Nucl. Phys. B302(1988)1;
\item{[17]} P. van Baal, Phys. Lett. B224(1989)397
\end